# Agile Collaboration for Distributed Teams

Fabio Calefato and Christof Ebert

**From the Editor**

Today, software engineering is characterized by two strong trends: agile and distributed. Both together are increasingly demanded and challenge teams and projects due to lack of discipline, insufficient transparency, agile "ping-pong," and thus overheads and rework. Authors Fabio Calefato and I describe current technologies and tools for agile collaboration. I look forward to hearing from both readers and prospective department authors about this article and the technologies you want to know more about. —*Christof Ebert*

**DISTRIBUTED AGILE TEAMS** traditionally have relied on developing custom interpretations of agile practices as well as on adopting an ever-growing plethora of tools.[1] However, this does not scale anymore to fast deliveries and an increasing need for governance, quality, and cost-efficiency. We present here some of the most successful tools used by companies to support collaboration in distributed agile teams. Albeit our review is not exhaustive, it nonetheless provides an up-to-date overview of current technologies. It is based on empirical studies from recent International Conference on Global Software Engineering meetings (see "International Conference on Global Software Engineering") as well as consulting with major software companies.

## Collaborative Tools for Distributed Agile Teams

Collaborative tools can be broadly grouped along three dimensions (see Figure 1):

- *Communication*. Tools under this category ensure understanding and exchange among distributed project stakeholders and engineers. Examples include instant messaging and VoIP tools, conferencing systems, and communication platforms that integrate collaborative features such as screen sharing and a calendar.
- *Workspace*. A distributed project requires technical support to manage a multitude of interdependent work products, with complex and evolving functionality, designs, and architectures. Examples of such technologies include design and modeling tools, project and software configuration management products, issue tracking and knowledge management systems, build and test environments, and package managers.
- *Lifecycle*. Software tends to evolve over time, with a longer

lifetime creating different versions or variants of the system, each of which must be managed until its end of life. Examples include some of the abovementioned technologies, such as traceability and test configurators for regressions, or application lifecycle management (ALM)/product lifecycle management (PLM), service request management, and DevOps and over-the-air delivery technologies.

Next, we review each category, with a focus on collaboration technologies. We argue that these tools are of paramount importance for distributed agile teams as they not only enable collaboration but also act as *hubs* through which relevant pieces of information flow,[2] thus helping team members build and maintain up to date their *situational awareness*—the state of mind where a person is aware of the elements in their immediate environments—about coworkers, tasks, and artifacts.[3]

## Communication

Software engineers are used to relying on a wide array of communication technologies, both asynchronous (email, forums) and synchronous (chat, audio conferencing, videoconferencing), when direct communication is not an option. Despite the steady technological advance, finding a solution that works is still a challenging task, both technically (firewalls, company restrictions) and economically (interoperability of preexisting solutions), even for small distributed teams. Table 1 highlights a selection of modern communication platforms along with their most notable features.

### INTERNATIONAL CONFERENCE ON GLOBAL SOFTWARE ENGINEERING

The annual International Conference on Global Software Engineering (ICGSE) brings together industry and research, providing the leading forum for addressing topics such as how to make distributed teams more effective and efficient and how to cope with challenges created by such distributed teams, such as different methods and tools.

ICGSE 2019 will take place 25–26 May in Montréal, Canada, colocated with the International Conference on Software Engineering. Join the conference and learn how to overcome challenges in agile and distributed software projects.

For more information, visit www.icgse.org.

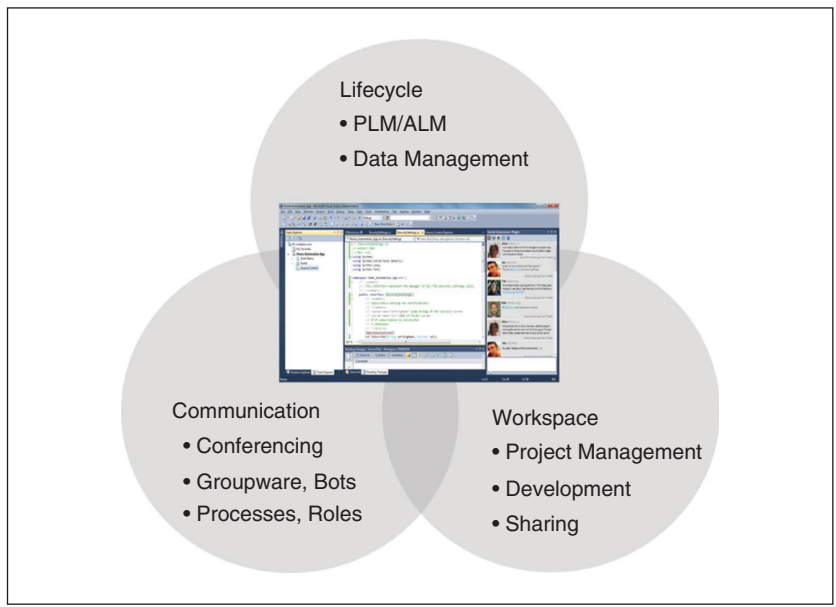

**FIGURE 1.** The collaboration infrastructure.

### Collaboration Platforms

There is a recent trend for distributed groups to collaborate using communication platforms that allow them to create a project workspace as a central hub where all intrateam conversations can be arranged in separate channels/rooms. In this category, the king of the hill is Slack (slack.com), whose enormous success has shaped a series of similar tools such as Microsoft Teams (products.office.com/en-us/microsoft-teams), a serious contender for the throne, especially for teams already using Microsoft solutions such as Skype for Business (www.skype.com/en/business) and Google Hangouts (hangouts.google.com), which has been repurposed from a casual chat

Table 1. Communication tools with native features.

| Tool | Communication channels | Searchable history | Audio/video calls | Videoconferencing | Recording | Dial-in | Screen sharing | File sharing | Calendar | Integrations | Mobile apps | License model |
|---|---|---|---|---|---|---|---|---|---|---|---|---|
| Gitter (gitter.im) | Yes | Yes | No | No | No | No | No | Only as links to GitHub | No | Yes (API) | iOS, Android | Open source |
| Google Hangouts (hangouts.google.com) | Yes | Yes | Yes | Yes (up to 50) | Yes | Yes | Yes | Yes | Yes (via Google Calendar) | Yes (no API) | iOS, Android | Freemium |
| GoToMeeting (www.gotomeeting.com) | No | No | Yes | Yes (up to 100) | Yes | Yes | Yes | Yes | Yes | Yes (API) | iOS, Android | Paid |
| Microsoft Teams (products.office.com/en-us/microsoft-teams) | Yes | Yes | Yes | Yes (up to 250) | Yes | Yes | Yes | Yes | Yes | Yes (API) | iOS, Android | Freemium |
| Skype for Business (www.skype.com/en/business) | Yes | Yes | Yes | Yes (up to 250) | Yes | Yes | Yes | Yes | Yes (via Outlook) | Yes (API) | iOS, Android | Freemium |
| Slack (slack.com) | Yes | Yes | Yes | Yes (up to 15) | No | No | Yes | Yes | No | Yes (API) | iOS, Android | Freemium |
| WebEx (www.webex.com) | No | No | Yes | Yes (up to 75) | Yes | Yes | Yes | Yes | Yes (via Google Calendar) | Yes (API) | iOS, Android | Paid |

tool to an enterprise communication solution.

On all these platforms, the interaction can be written and audio/video, allowing users to switch seamlessly from sync to async, and vice versa, thus replacing intrateam email communication. The history of events is also logged and searchable online, preventing loss of knowledge or silos. These platforms provide native support for basic collaborative features such as calendar, file, and screen sharing; in addition, they allow for the lightweight interconnection of countless services for software development, with which engineers can interact by just typing in commands from a familiar user interface.[4] This form of conversation-driven development, or *ChatOps*, has been particularly beneficial to foster DevOps practices in distributed projects by making the execution of operations tasks more accessible to developers.

### Web Solutions
The classic alternative to the aforementioned new communication platform is using web conferencing solutions such as Cisco WebEx (www.webex.com) and GoToMeeting (www.gotomeeting.com). The most notable difference between these two types of communication solutions is that the web conferencing tools do not provide a workspace for chat communication arranged in channels. Also, they typically offer only paid layers, whereas Slack, Microsoft Teams, and Google Hangouts are freemium products whose free layer is particularly appealing to small teams who do not need group conferencing.

Another communication tool that has been gaining momentum in this post-email era is Gitter (gitter.im), a lean, open source communication platform that is particularly appealing to distributed projects using GitHub because it allows teammates to discuss around artifacts (issues, commits, pull requests) by just adding direct links to the repository. However, we point out that Gitter only supports text-based interaction.

All these communication products offer full-featured mobile apps, which allow remote team members to collaborate smoothly also on the go.

## Workspace
Agile project development elevates the idea of self-organizing and self-managing teams, requiring software engineers to embrace behaviors of flexibility, empowerment, trust, and collaboration. As such, adequate tool support is necessary for coordinating development activities in agile teams, especially when distributed.[7]

### Project Workspace
The fundamental activity of any software team is developing source code. GitHub (github.com) and GitLab (gitlab.com) are platforms, often referred to as *Collaborative Development Environments*,[5] that provide a project workspace that increases developers' productivity and comfort by integrating in one place a standardized toolset consisting of a version control system, such as Git, to share software artifacts in a controlled manner an issue tracking system, i.e., a database to manage defect reports and change requests, and a content management system to store explicit knowledge.

### Requirements and Design
In agile projects, requirements are expressed using natural language as user stories that are then stored in an issue tracking system. While issue trackers are typically integrated into the project workspace thanks to tools like GitHub and GitLab mentioned previously, there are also specific products such as Pivotal Tracker (www.pivotaltracker.com) and Atlassian Jira (www.jira.com), which support task management, through Kanban or Scrum boards, and analytics for performance evaluation with agile metrics (e.g., burndown and velocity charts).

Collaboration in software teams is not only about sharing files but also content, as in the case of design models.[6] A few web-based collaborative diagramming tools for UML are available, such as Creately (creately.com), which integrates natively with Jira, Cacoo (cacoo.com), and GenMyModel (www.genmymodel.com).

### Knowledge Centers
Distributed projects need a content management system to share explicit knowledge such as internal documents, standards, and best practices. Typically, teams use wikis because they are easy to edit, automatically shared, and often integrated within the project workspace as in GitHub and GitLab. However, modern teams now also rely on Q&A platforms such as Stack Overflow Enterprise to store and make existing team knowledge easier to search. Free, open source Q&A forums also exist, such as Question2Answer (www.question2answer.org), Talkyard (www.talkyard.io), and Scoold (scoold.com).

### Building and Testing
Distributed projects have a great need for secure, remote repositories and build management. The selection of build tools obviously depends on the programming language. Java

**COLLABORATIVE ALM/PLM CASE STUDY**

Companies increasingly realize that collaboratively working at a higher level of abstraction and automation will improve productivity and quality. Model-based development plays a pivotal role in this evolution. The companies we work with share the same goals: master complexity, improve product quality, shorten development time, and plan functions in different variations and versions for better reuse. The biggest challenges they see are their own learning curves and keeping consistency across features and products. Systems engineering is growing fast to have top-down consistency across artifacts, components, and evolving functionality. To get insight into achieved benefits, we followed through like in other improvement projects by looking into performance results from real projects as well as some process-related aspects, such as knowledge utilization. We found better quality, reduced cycle time, more flexibility, lower overheads by improved communication, increasing alignment of processes and tools, and faster ramp-up time and competence management across distributed teams. The business case is clear: with a degree of at least 30% for consistent modeling, error rates and development costs are reduced significantly. Improved visibility and aligned terminologies and roles are low-hanging fruit and deliver big gains, as they facilitate a borderless solution building.

also fosters DevOps collaboration by blurring, and eventually eliminating, the distinction between the roles of development and operations, with developers able to manage and provision servers and containers through machine-readable definition files (or *infrastructure as code*) across the entire project lifecycle.

### Lifecycle

The growing complexity of modern software is directly proportional to the size of the product stack that companies need to use. With increasing delivery speed and necessary flexibility, collaborative agile lifecycle management is mandatory for development teams. Daunting as it may appear, interacting with appropriate tools across the lifecycle is vital because they provide distributed agile teams with a comprehensive solution to perform and collaborative tasks.

### ALM/PLM

PLM and ALM provide support across the lifecycle management of a product. (See "Collaborative ALM/PLM Case Study.") PLM considers products with many different artifacts, such as critical mechatronic systems and embedded electronics. ALM focuses primarily on IT and software applications without the needs of critical systems such as functional safety. Few suites support real collaboration with simultaneous changes and automatically allocate changes to impacted artifacts based on the changes made.

projects use Maven (maven.apache.org) or Gradle (gradle.org) for compiling and resolving external dependencies, whereas Python and JavaScript projects rely, respectively, on pip and npm as package managers.

According to the Agile Manifesto, at the heart of agile development are four actions: collaborate, deliver, reflect, and improve. Testing is all about reflection and improvement. Therefore, testing also makes up the backbone of the agile methodology. Also in this case, tool selection depends on the programming language. JUnit is the de facto standard for executing unit tests in Java projects. Projects building on JavaScript frameworks often use Jasmine (jasmine.github.io) for writing behavioral-driven test cases and Karma (karma-runner.github.io) as a test runner. Web applications also use Selenium (www.seleniumhq.org) as a web driver to automate testing across many browsers.

While presented as separated, building and testing are now often performed together using tools for continuous integration (CI), which, upon code changes pushed to the main line of development, automate the execution of scripts for building software, invoking testing frameworks and deploying/delivering successful releases to the appropriate environment (staging or production). Travis CI (travis-ci.org), Jenkins (jenkins.io), and CircleCI (circleci.com) are among the most used CI solutions.

Supporting CI in agile projects not only grants workflow automation but

Current popular solutions include Microsoft Visual Studio Team Foundation Server (www.visualstudio.com/tfs), an integrated collaboration suite of developer tools, build system, and version control; IBM Rational Collaborative Lifecycle Management (https://

DRAFT

DRAFT

www.ibm.com/us-en/marketplace/application-lifecycle-management), with a highly integrated suite of tools such as Team Concert, DOORS NG, and Quality Manager; Siemens Teamcenter Polarion (polarion.plm.automation.siemens.com), an integrated platform to automate the development processes across various projects; Micro Focus Connect (formerly Borland Connect, www.microfocus.com/products), which builds upon design and modeling support with collaborative deliveries; and Versionone (www.versionone.com), built around an agile project management solution and development software platform.

When selecting a lifecycle management solution, be aware of the complete lifecycle cost and potential lock-in mechanisms that make it hard to escape or change later. Highly integrated ALM/PLM suites, such as IBM Rational, offer deeply embedded trigger and consistency mechanisms, but at the cost to lock in users with a high effort to switch to another environment later. Traditional agile environments such as Polarion are eventually integrated into bigger suites, reducing the chances for small setups and affordable license schemes. Therefore, we recommend federated approaches that allow flexible evolution in line with the growth of your business. Always consider the exit strategy with export mechanisms to other tools, such as Requirements Interchange Format in the domain of requirements tools.

### Quality Control

Global software development challenges traditional quality control and asks for new solutions. Given the global competition, quality must be good enough and proven for any component and for the entire system. Distributed ownership of these various software components does not allow close teamwork toward continuous builds or peer reviews. Often, the owners of interacting components know each other only from phone conversations but still must assure the right level of quality control. As a first step, of any quality control activity, one must define the quality levels to be achieved. In global development projects, this is often done by means of a service level agreement or by phase-end or hand-over criteria. These targets must be measurable.

Independent of what global collaboration model or contract model is established with suppliers, it is key to set the right targets and set them as performance indicators for R&D the management in each location. Technologies supporting verification and validation include static analysis and continuous unit test (based on test-driven development), which can be directly included in continuous development workbenches. Code complete in distributed agile teams should be based on two trigger criteria: no major warnings from static analysis and 100% C0 coverage. Many tools provide support for verification and validation, so we will not go into more details here.[4,6]

Here is a warning on work allocation and ownership. Shifting verification activities to low-cost countries is highly inefficient. Tasks are often overly fragmented, and the quality control activities are handled with poor results due to lack of knowledge. In the end, each delivery must be checked twice, at the time it is shipped to a low-cost country and then again backward. All this costs time and money, and it demotivates engineers on both sides, as it always

> At the heart of agile development are four actions: collaborate, deliver, reflect, and improve.

ends up in ping-pong. As said before, we strongly recommend building teams preferably in one place and assigning them ownership for a work product including functionality and quality. Such teams should operate globally according to needs and skills availability but not be internally split into first- and second-class engineering tasks.

### Are Collaborative Technologies Necessary?

Collaboration is strategic to distributed agile teams, and appropriate tool support is the only way to do this efficiently. Still, consolidating a product stack is a complex endeavor, which entails evaluating alternative solutions and adapting to changes, and, therefore, it is best performed as a stepwise process.

However, activities such as project management, predevelopment, and product engineering are rarely integrated well due to the diversity of stakeholders with individual knowledge about projects, products, and processes. Therefore, engineering results such as specifications, documentation, and test cases are

## ABOUT THE AUTHORS

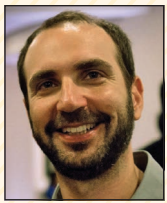

**FABIO CALEFATO** is an assistant professor of computer science at the University of Bari, Italy. He is the general chair of ICGSE 2019. Contact him at fabio.calefato@uniba.it.

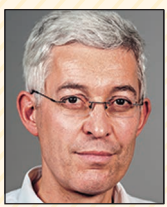

**CHRISTOF EBERT** is the managing director of Vector Consulting Services. He is on the *IEEE Software* editorial board and teaches at the University of Stuttgart and the Sorbonne in Paris. Contact him at christof.ebert@vector.com.

inconsistent, and items like signals and parameters are arbitrarily labeled. Changes create a lot of extra work to make sure that nothing is overlooked. Reuse is hardly possible due to the many heterogeneous contents. This pattern is amplified when collaboration across supplier networks and complex workflows take place. Distributed agile teams and sourcing without adequate tools and IT infrastructure will immediately cause problems and high overheads. Collaboration across enterprises and distributed teams often leads to fragmented processes and toolchains with heterogeneous interfaces and redundant and inconsistent data management. This results in insufficient transparency.

The move to global agile teams is a useful catalyst to incrementally clean up such legacy tools inventory. Effective tool support for collaboration is a strategic initiative for any company with distributed resources no matter if the strategy involves offshore development, outsourcing, or supplier networks. Software needs to be shared, and appropriate tool support is the only way to do this efficiently, consistently, and securely.

What is next in agile collaboration? Knowledge management must be better linked to business. Aligned business objectives and metrics must guide and monitor the development processes, product lines, and project teams. Take as an example a mobile phone or game design with lots of embedded software. Being a commodity, business-oriented targets cover return rates or brand loyalty. Defects increase return rate and reduce brand loyalty with devastating business impacts. Looking at projects, products, and processes will improve the design away from an overly narrow focus on manufacturing aspects toward usability engineering. Knowledge and experience from past projects will be embedded into the underlying design processes. We stress the need for adequate knowledge management as a basis for success in product and solution development, an aspect going well beyond most collaborative approaches of today. Yet, personal contact will always be necessary to provide context and analysis. Collaborative technologies should, therefore, facilitate interpersonal communication and sharing of artifacts.